\documentclass[12pt,a4paper,twoside,english,aps,reprint,showpacs]{revtex4-1}
\usepackage{lmodern}

\usepackage[T1]{fontenc}
\usepackage[latin9]{inputenc}
\usepackage{fancyhdr}
\usepackage{babel}
\usepackage{float}
\usepackage{textcomp}
\usepackage{ifsym}
\usepackage{mathrsfs}
\usepackage{amsmath}
\usepackage{amssymb}
\usepackage{graphicx}
\usepackage{esint}
\usepackage[unicode=true,
 bookmarks=true,bookmarksnumbered=true,bookmarksopen=true,bookmarksopenlevel=1,
 breaklinks=false,pdfborder={0 0 0},backref=false,colorlinks=false]
 {hyperref}
\hypersetup{pdftitle={Your title},
 pdfauthor={Your name},
 pdfpagelayout=OneColumn, pdfnewwindow=true, pdfstartview=XYZ, plainpages=false}

\makeatletter

\pdfpageheight\paperheight
\pdfpagewidth\paperwidth

\usepackage{amsmath}    
\usepackage{graphicx}   
\usepackage{verbatim}   
\usepackage{color}      
\usepackage{subfigure}  
\usepackage{hyperref}   
\usepackage{dcolumn}
\usepackage{bm}

\makeatother

\begin{document}
\global\long\def\mb#1{\mathbf{#1}}

\global\long\def\mr#1{\mathrm{#1}}

\global\long\def\dthree#1{\mathrm{d}^{3}\mathbf{#1}}

\global\long\def\cop#1{\hat{#1}^{+}}

\global\long\def\aop#1{\hat{#1}}

\global\long\def\Ket#1{\left|#1\right\rangle }

\global\long\def\Bra#1{\left\langle #1\right|}

\title{The effect of anisotropic exchange interactions and short-range phenomena
on superfluidity in a homogeneous dipolar Fermi gas}

\date{\today}

\author{I. Corro }

\author{A. M. Martin}

\affiliation{School of Physics, University of Melbourne, Parkville, Victoria 3010,
Australia.}
\begin{abstract}
We develop a simple numerical method that allows us to calculate the
Bardeen-Cooper-Schriefer (BCS) superfluid transition temperature ($T_{c}$)
precisely for any interaction potential. We apply it to a polarised,
ultracold Fermi gas with long-range, anisotropic, dipolar interactions
and include the effects of anisotropic exchange interactions. We pay
particular attention to the short-range behaviour of dipolar gasses
and re-examine current renormalisation methods. In particular, we
find that dimerisation of both atoms and molecules significantly hampers
the formation of a superfluid. The end result is that at high density/interaction
strengths, we find $T_{c}$ is orders of magnitude lower than previous
calculations.
\end{abstract}


\maketitle
A great deal of interest in dipolar Fermi gasses has been generated
due to their long range interactions, which lead to many novel effects
such as p-wave superfluidity \citep{2002-PRA_Baranov-Shly_BCSDipRenorm,2004-PRL_Bara-Dob-Lewen_BCSDipTrappedPwave,2008-PhysRep_Baranov_DipGasRev,2008-PRL_BruunTaylor_2DDipFermGasPhases},
topological superfluidity in 2D systems \citep{2009-PRL_Cooper-Shly_BCSof2DIntTaylor,2011-PRA_Levinsen-Shly_2DIntTaylorAndSuperFl},
anisotropic and many body effects on the Fermi liquid properties \citep{2012-ChemRe_Bar-Dal-Pup-Zoller_BigDipolarReview,2015-PRA_Krieg-Kopietz_SecondOrderDipoleCalc},
the tailoring of novel interaction potentials \citep{2007-PRL_Buchler-Zoller_InterTaylor,2007-PRA_Micheli-Zoller_InterTaylor},
and superfluidity in bilayers \citep{2010-PRL_Shlyapnikov_BilayerFermSuperfluid,2011-PRA_BaranovZoller_BilayerDipFerm}.

This rich selection of interesting phenomena has lead to a large effort
from many groups to trap and cool a dipolar gas to degeneracy. Many
highly successful experiments have resulted from this effort, which
have concentrated on both molecular \citep{2010-PRL_JILA_HyperFineGState,2010-Nature_JILA_UColdDipFermiGas,2010-Science_JILA_ChemReacControlKRb,2011-NatPhys_JILA_ChemStereoDyn,2012-PRL_Harvard_NaK-AbsGS,2012-PRL_JILA_PolarizabilityOfKRb,2015-PRL_Harvard_NaKGstate}
and atomic, highly-magnetic dipolar gasses \citep{2012-PRL_Lu-Burdick-Lev_DyFermiGas,2014-PRL_Innsbruck_ErDeg,2014-Nature_Frisch_Qchaos,2014-PRA(R)_Stanford_FeshResInDy,2014-PRL_Aikawa_RelaxDynam,2014-Science_Innsbruck_FermiDeformation,2015-PRA_CNRS_DegenCr,2015-PRL_Stanford_FermSuppOfDipRelax}.
These experiments investigated features such as the precise control
of ultracold chemical reactions \citep{2010-Science_JILA_ChemReacControlKRb,2010-Nature_JILA_UColdDipFermiGas,2011-NatPhys_JILA_ChemStereoDyn},
quantum chaos in dipolar collisions \citep{2014-PRA(R)_Stanford_FeshResInDy,2014-Nature_Frisch_Qchaos},
anisotropic interaction effects in the Fermi surface \citep{2014-Science_Innsbruck_FermiDeformation},
and dipolar collisions \citep{2014-PRL_Aikawa_RelaxDynam}. As yet,
however, a dipolar Fermi superfluid has not been observed. 

Predictions for the superfluid transition temperature of a dipolar
Fermi gas ($T_{c}$) have been calculated in a number of works under
various conditions \citep{2002-PRA_Baranov-Shly_BCSDipRenorm,2008-PRL_BruunTaylor_2DDipFermGasPhases,2009-PRL_Cooper-Shly_BCSof2DIntTaylor,2010-PRA_Zhao-Pu_BCSDipGas,2010-PRL_Shlyapnikov_BilayerFermSuperfluid,2011-PRA_BaranovZoller_BilayerDipFerm,2011-PRA_Sieb-Baran_2DDipFermGas,2011-PRA_Levinsen-Shly_2DIntTaylorAndSuperFl,2015-PRL_Beijing_WeylSuperInDipFerm}.
These works consider a dipolar Fermi gas within an idealised condensed
matter paradigm, which is usually applicable for thermodynamically
stable systems. However, an ultra-cold dilute gas is not stable. We
investigate the complications that this introduces and produce our
own predictions for Tc that differ significantly from these previous
works at high densities or interaction strengths. After deriving our
results, we compare our methodology with previous works in Section
\ref{sec:Conclusion}.

This paper will be set out as follows. In Section \ref{sec:dip-bond}
we review some important theoretical and experimental background involving
the stability of p-wave gasses, particularly the collapse into p-wave
dimers. We will then investigate how adding long-range interactions
affects these p-wave dimers and produces dipolar-interaction-dominated
bound states. In Section \ref{sec:dipbond_and_Tc} we see that these
dipolar bound states can have a large effect on the stability of the
gas and therefore also on $T_{c}$. A key result of this paper is
that we will show that in attempting to renormalise out the short
range behavior of a dipolar gas, previous works in this area have
ended up calculating a transition to tightly bound molecules, rather
than to a BCS superfluid. We will then suggest a remedy to this problem. 

Section \ref{sec:Numerical-method} describes a simple algorithm that
allows one to easily calculate the BCS transition temperature for
complicated potentials. Furthermore, it allows us to take into account
the effect of an anisotropic Fermi surface (in this case caused by
the anisotropic interaction potential of a polarised dipolar gas).
This is a general algorithm that can be applied to any system. In
Section \ref{sec:Numerical-Results} we apply the numerical method
from Section \ref{sec:Numerical-method} to the theoretical method
of Section \ref{sec:dipbond_and_Tc} to obtain prediction for $T_{c}$
at different experimental parameters. We find that in the region of
experimental interest (i.e., high interaction strength or high densities)
is where our results differ from previous theoretical works in this
area, giving us a $T_{c}$ which is much lower. In Section \ref{sec:Conclusion}
we will then compare our results to this previous theoretical work
in more detail with the aim of understanding how and why they differ.

\section{Understanding the role of dipolar bound states and the centrifugal
barrier.\label{sec:dip-bond}}

Dipolar gas experiments face a number of challenges, which include
dipolar spin-flip collisions in atoms \citep{2015-PRL_Stanford_FermSuppOfDipRelax},
chemical reactions in dipolar molecules \citep{2010-Nature_JILA_UColdDipFermiGas,2010-Science_JILA_ChemReacControlKRb,2011-NatPhys_JILA_ChemStereoDyn},
and long-lived scattering chain complexes \citep{2013-PRA_JILA-Que-Bohn_ScatChainComplex}.
Because s-wave interactions are not allowed between identical fermions,
all these effects can be protected against using the centrifugal barrier
in the p-wave (and higher) scattering channels to prevent two scattering
dipoles from coming into contact. The great effectiveness of the p-wave
barrier in protecting the gas from inelastic processes has been shown
in a series of experiments involving $\mr{^{40}K^{87}Rb}$ molecules
at JILA \citep{2010-Nature_JILA_UColdDipFermiGas}, which are prone
to react chemically (via $\mr{KRb+}$$\mr{KRb}\rightarrow$$\mr K_{2}+$$\mr{Rb_{2}}$),
as well as in Dy atoms, which undergo spin flip collisions \citep{2015-PRL_Stanford_FermSuppOfDipRelax}.

To help understand the properties of the centrifugal barrier in the
presence of dipolar interactions, we recall the results of Ref. \citep{2010-PRA_Que-Bohn_DipThreshScatMod}.
In that work, it was shown that the attribute of interest is the height,
$V_{b}$, of the barrier with the lowest maximum strength (which is
the $l=1$, $m=0$ barrier) and how this compares with the average
particle kinetic energy. It was also shown that the location and height
of this barrier varies as the interactions strength, $C_{dd}$, changes,
with $V_{b}$ decreasing as $C_{dd}$ increases, and finally it was
shown that for sufficiently strong $C_{dd}$, $V_{b}$ is completely
determined by $C_{dd}$ and is independent of the short range specifics
of the potential. The result is that within this regime \citep{2010-PRA_Que-Bohn_DipThreshScatMod}
\begin{equation}
r_{b}=\alpha\frac{C_{dd}m}{\hbar^{2}},\label{eq:rb}
\end{equation}
where $\alpha=0.6$, and $m$ is the mass of the particles. We will
use units such that $r_{b}=1$, $\hbar=1$, $k_{B}=1$, and $m=1$,
giving units of energy $E_{D}=\hbar^{2}/mr_{b}^{2}=\hbar^{6}/m^{3}C_{dd}^{2}\alpha^{2}$.
In these units, $C_{dd}=1/\alpha$ and $V_{b}=2/3$ are both constants.
All our equations will reduce down to just two dimensionless parameters:
the dimensionless temperature $\mathfrak{\tau}=T/E\propto T/C_{dd}^{2}$,
and the dimensionless average distance between atoms $\lambda=\rho^{-1/3}/r_{b}\propto C_{dd}\,^{-1}$,
where $\rho=k_{F}{}^{3}/6\pi^{2}$ is the density, and $k_{F}$ is
the Fermi surface (which gives $\lambda\approx3.9/r_{b}k_{F}$). Ref.
\citep{2010-PRA_Que-Bohn_DipThreshScatMod} shows, as is verified
by experiments \citep{2010-Nature_JILA_UColdDipFermiGas}, that as
the average particle kinetic energy approaches $V_{b}$, the rate
of barrier transmission nears unity, and quenching rates become unacceptably
high. We note, as a benchmark for this effect, that the Fermi energy
intersects $V_{b}$ at $\lambda\approx3.38$.

An important conclusion from Eq.~(\ref{eq:rb}) is that in a gas
with polarised dipolar interactions, the centrifugal barrier is much
farther away from $r=0$ than a gas with short range interactions.
Because dipole interactions are strongly attractive in the $l=1$,
$m=0$ subspace, they can lead to dipolar bonding between dipoles
\citep{2011-PRL_WangIncGreene_3BodyDipFermion,2012-PRA_WangGreene_BoundAndScatPropOfTwoDip}.
These dipole-induced, p-wave dimers will be of the size of the centrifugal
barrier and will therefore be much larger than the p-wave bound states
found in typical gasses with short-range interactions. The formation
of p-wave dimers is significant because they are well known to be
unstable from investigations in systems of identical fermions with
short-range interactions near a p-wave resonance. These systems were
first shown to be experimentally unstable \citep{PWE_2003-PRL_Regal-Tick-Bohn_K,PWE_2004-PRA_Ticknor-Bohn-Jin,PWE_2004-PRAR_ENS_6LiRes,PWE_2005-PRA_MIT_Li6-s-p-wave,PWE_2005-PRL_Zurich_2Dpwavein40K,PWE_2007-PRL_Gaeb-Bohn-Jin_FeshMol-K,PWE_2008-PRA_ARC_FeshMolBindEn-Li1,PWE_2008-PRL_Tokyo_6LiMolCollProp,PWE_2010-PRA_Maier-Zimm_RFspecOn6LiMol,PWE_2013-PRA_Tokyo_6LiScatParam},
then it was then shown that the major cause of this instability could
be attributed to the fact that p-wave dimers, which are formed rapidly
via three-body interactions when near the resonance, are unstable
to collisional relaxation and decay much faster than they can thermalise
\citep{2007-PRL_Levinsen-Coop-Gurarie_StrongResPWaveSF,2008-PRA_Dinc-Esry-Greene_UcoldAtomMolColl,2008-PRA_Levinsen-Gurarie_StabilityOfFermiGasesPwave,2008-PRA_Lasinio-Pric-Castin_3ferm,PWE_2008-PRA_ARC_FeshMolBindEn-Li1}.
It has been shown that this behavior is a general property of identical
fermions, and not of any property of the particular species being
investigated \citep{2007-PRL_Levinsen-Coop-Gurarie_StrongResPWaveSF}.

\begin{figure}[h]
\noindent \centering{}\includegraphics[width=8.6cm]{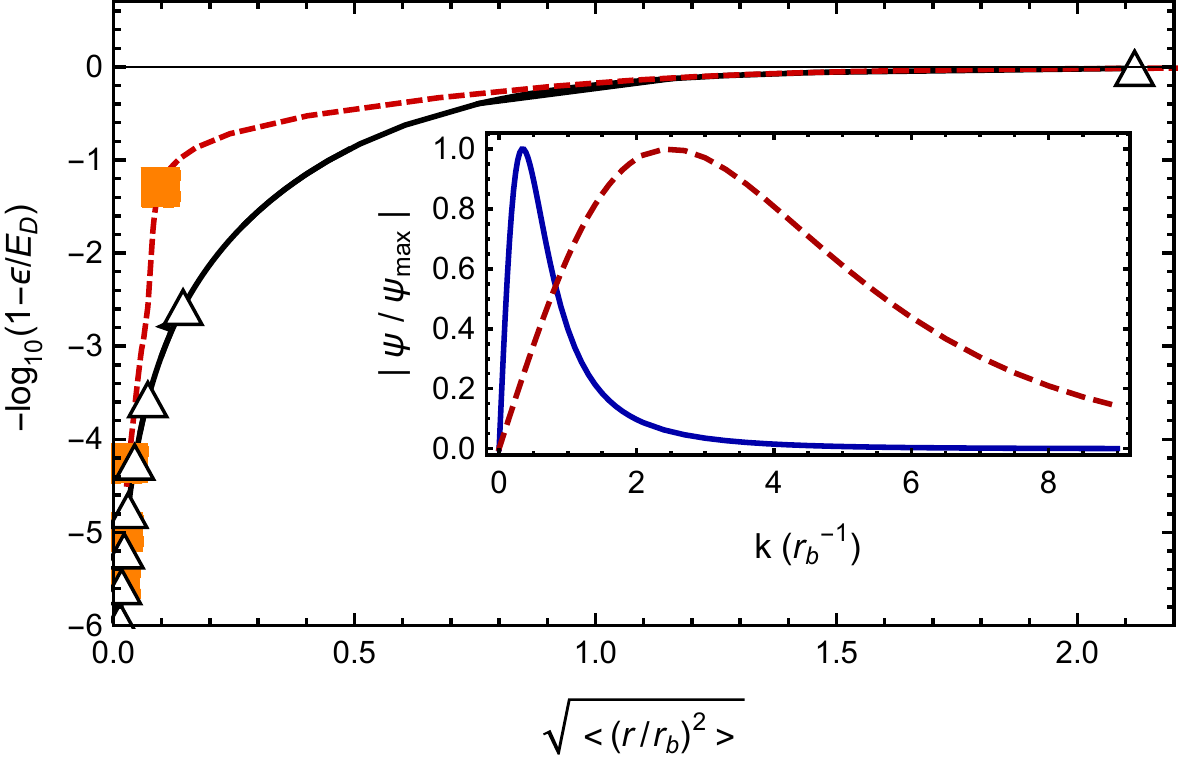}\protect\caption{\label{fig:dip_boundstate}The bound state energies vs the root mean
square size of the corresponding wavefunctions for a dipolar potential
with a cutoff at $R_{cut}$. The triangles and squares represent bound
states at $R_{cut}=0.01093r_{b}$, just before the largest bound state
disappears. As $R_{cut}$ varies, all the squares move along the exact
same dashed line, and all the triangles move along the same solid
line. The triangle states are $l=1$ and 3 dominated. The square states
are $l=5$ and 7 dominated. Interestingly, the bound states are all
evenly spaced along each line. For example, depicted here the $7^{{\rm th}}$
triangle, sitting at $2.1r_{b}$, is about to disappear and the $6^{{\rm th}}$
triangle sits at $0.15r_{b}$. If we continue to increase $R_{cut}$
until the $6^{{\rm th}}$ triangle is about to disappear, we find
that the $6^{{\rm th}}$ triangle will be at $2.1r_{b}$ and the $5^{{\rm th}}$
triangle at $0.15r_{b}$, almost exactly where the $6^{{\rm th}}$
triangle is now. Hence we can infer that the possible sizes of the
largest bound state should sit within the range $0.15r_{b}$ to $2.1r_{b}$.
Inset: the wavefunction in momentum space of the eigenstate that is
on resonance (filled line), and the next shallowest bound state (dashed
line). Both are the $l=1$, $m=0$ component. The filled line is therefore
the largest possible dimer (occuring when the potential is on resonance).
The dashed line is the smallest possible size for the shallowest dimer
(occurs when the largest bound state disappears).}
\end{figure}

In Fig.~\ref{fig:dip_boundstate}, we investigate the behavior of
a p-wave dimer of two particles with strong dipolar interaction. The
exact details of the interaction potential is species dependent; however,
for particles with strong dipolar interactions, which are of interest
to us, the largest bound states are dominated by the bare dipolar
interaction. We have solved the Schr\"{o}dinger equation using the
finite element method \citep{Book_2002_Finite-Element-In-QM} for
a bare dipolar interaction with a cutoff, $R_{cut}$, and plotted
the energy and the root-mean-squared widths of the eigenstates. The
short-range behavior is unknown, so we have plotted these eigenstates
for a continuously varying value of $R_{cut}$ (filled and dashed
lines), showing the effect that different short-range behaviour can
have on the bound states of a dipolar-long-range-dominated interaction
potential. The figure shows the typical behavior of the bound states,
which are evenly spaced and increase in size as $R_{cut}$ increases.
Just before the energy of a bound state crosses zero, the bound state
is at its largest, with a root-mean-square extent of $2.1r_{b}$.
Just after the energy goes above zero, the bound state disappears
and the next bound state becomes the largest, which is always sitting
at around $0.15r_{b}$. We can therefore conclude that no matter the
short-range specifics, the root mean squared size of the largest bound
state will be within the range $0.15r_{b}$ and $2.1r_{b}$.

Three-body dimerisation rates for identical dipolar fermions have
been explicitly calculated in \citep{2011-PRL_WangIncGreene_3BodyDipFermion},
but not relaxation rates. In the s-wave case, the increasing size
of a dimer leads to stability against inelastic relaxation collisions
as the distance between the shallowest dimer and the deeply-bound
states increases \citep{2007-PRL_Levinsen-Coop-Gurarie_StrongResPWaveSF,2003-PRA_Petrov_3BodProbFermGas,2004-PRL_Pet-Sal-Shly_WeakDimersOfFermAtoms,2005-PRA_Pet-Sal_Shly_WeakDimersOfFermAtoms,2001-PRL_EsryGreeneSuno_3BodyRecombThreshLaw,2003-PRL_SunEsGreene_RecombOf3Fermions}.
In the dipolar case, the sizes of the deeply bound states increase
along with the size of the shallowest dimer, negating that protection.
Fig.~\ref{fig:dip_boundstate} shows that once a weakly bound dimer
of two dipoles is formed, there is cascade of nearby tighter bound
states that the dimer can easily decay into. These dimers can then
be expected to be unstable just as for the short-range case.

\section{The effect of dipolar bonding on $T_{c}$\label{sec:dipbond_and_Tc}}

Here we will show that the formation of these p-wave dimers is also
extremely important for the renormalisation of the BCS equations.
The BCS equations find the minimum of the free energy. It is well
known from the BCS-BEC crossover problem \citep{1985_Nozieres_BCSBEC}
that if the full potential is considered, the BCS equations will simply
converge to the tightest bound state and form tightly bound bosons.
Ultra-cold gasses are meta-stable, and we are not interested in the
absolute ground state. For this reason, the BCS equations are usually
renormalised using the method of Randiera et. al \citep{1989-PRL_Randiera,1990-PRB_Randiera}.
The purpose being to remove the short range behavior, but capture
accurately the long range scattering properties of the atoms involved.
A similar method must be implemented for the case we are interested
in: a meta-stable gas of dipoles that sit outside each others centrifugal
barriers and interact only via the universal, long-range dipolar interaction. 

Predictions for $T_{c}$ of a dipolar Fermi gas under various conditions
have been calculated in a number of works \citep{2002-PRA_Baranov-Shly_BCSDipRenorm,2008-PRL_BruunTaylor_2DDipFermGasPhases,2009-PRL_Cooper-Shly_BCSof2DIntTaylor,2010-PRA_Zhao-Pu_BCSDipGas,2010-PRL_Shlyapnikov_BilayerFermSuperfluid,2011-PRA_BaranovZoller_BilayerDipFerm,2011-PRA_Sieb-Baran_2DDipFermGas,2011-PRA_Levinsen-Shly_2DIntTaylorAndSuperFl,2015-PRL_Beijing_WeylSuperInDipFerm}.
The most common method used is to introduce the renormalised equations
of Randiera et.al., except instead of replacing the T matrix with
its long-range behavior via a first order expansion in $k$ (i.e.,
$4\pi\hbar^{2}a/m$), the Born approximation is used to replace the
T-matrix by the bare potential. However, it is not clear that the
short-range behavior is removed with this technique. Indeed, the bare
dipolar potential contains all powers of $k$, meaning short-range
behavior is merely modified, not removed. In fact, if we examine Fig.~2
in Ref. \citep{2002-PRA_Baranov-Shly_BCSDipRenorm} or Fig.~3(b)
in Ref. \citep{2015-PRL_Beijing_WeylSuperInDipFerm}, we can see that,
for sufficiently large $k_{F}$, the gap plotted in those works is
the same size and shape as the two-body bound state plotted in the
inset in Fig.~\ref{fig:dip_boundstate} in this work. For $k_{F}\approx0.5/r_{b}$
($\lambda=7.8$) the gap in those works is the same size and shape
as the largest possible size for the shallowest dimer (filled line
in inset in Fig.~\ref{fig:dip_boundstate}). For $k_{F}\approx2.5/r_{b}$
($\lambda=1.56$) the gap in those works is the same size and shape
as the smallest possible size for the shallowest dimer (dashed line
in inset in Fig.~\ref{fig:dip_boundstate}). A true BCS pairing wavefunction
should be many times the size of the inter-particle spacing, not the
size of a bound state. This suggests that the BCS equations, renormalised
in this way, have simply picked up the tightly bound p-wave dimers
that we expect to be unstable. In other words, at high densities/interaction
strengths, these papers are calculating the transition to a tightly
bound BEC state rather than the desired BCS state.

In this paper we produce our own predictions for a 3D, polarized,
dipolar Fermi gas in a way that deals with these issues. We first
consider just the case of the KRb experiment at JILA, the solution
of which will turn out to be relevant to all systems discussed above.
We require a methodology which can describe a situation where the
gas is in quasi-stable equilibrium with molecules sitting outside
each others centrifugal barriers, and as soon as the particles tunnel,
they are almost guaranteed to be lost from the trap. We know that
the solutions should be independent of the short range details of
the molecules and should depend only on the $r^{-3}$ dipolar interaction.
However, if we were to simply use the bare dipolar interaction, the
BCS equations would only pick up the tightly-bound states that sit
well within the centrifugal barrier and do not represent the meta-stable
equilibrium that is desired.

The problem is that the dipolar interaction becomes very strong inside
the centrifugal barrier, but in the KRb experiments at JILA, the molecules
in quasi-stable equilibrium never ``feel'' that part of the potential,
and any particles that do venture within each others centrifugal barrier
undergo inelastic collisions with close to unit probability\citep{2010-PRA_Que-Bohn_DipThreshScatMod,2010-Nature_JILA_UColdDipFermiGas}
and consequently cannot contribute to the superfluid. We therefore
use the following \emph{effective }potential, which represents the
anisotropic interaction between \emph{meta-stable} dipoles that are
polarized in the $\hat{\mb z}$ direction.
\begin{equation}
V_{\mr{dd}}^{\mr{(eff)}}(\mb r)=\begin{cases}
C_{dd}\frac{1}{r^{3}}\left(1-3\cos^{2}(\theta_{r})\right) & r>r_{b}\\
0 & r<r_{b}
\end{cases},\label{eq:Veff_r}
\end{equation}
where $C_{dd}$ is the interaction coupling constant. $\theta_{r}$
is the angle between $\mb r$ and the $\hat{\mb z}$ axis. Eq.~(\ref{eq:Veff_r})
is just the usual dipolar potential \citep{2008-PRL_Ticknor_DipScatMultipleShortRange},
except cutoff at $r=r_{b}$. By using Eq.~(\ref{eq:Veff_r}), we
\emph{can} minimise the free energy without picking up the undesired
bound states, and the dipoles will feel the full dipolar potential
outside the centrifugal barrier, but not inside, exactly as the molecules
in the JILA experiments do. Furthermore, Eq.~(\ref{eq:Veff_r}) reproduces
the desired universal, dipolar-scattering amplitudes \citep{2008-PRL_Ticknor_DipScatMultipleShortRange,2009-NJP_Bohn-Ticknor_DipScatUnivers},
which means that provided the short range contribution to the potential
does not put the system on resonance, Eq. (2) contains the key properties
desired from a more conventional renormalisation method (see the discussion
in Section V).

\section{Numerical method \label{sec:Numerical-method}}

We use the following method to find $T_{c}$, which is generally applicable
to any potential, and can easily include the effects of the anisotropic
exchange interactions. First recall the standard result that the BCS
equations are equivalent to minimising the BCS free energy ($F$)
\citep{Book_1966_DeGennes}. This free energy can be written in terms
of the gap, $\Delta(\mb k)$, the potential in momentum space, $V(\mb k)$,
and the non-superfluid part of the quasi-particle energy, $\kappa(\mb k)=\hbar^{2}k^{2}/2m-\mu+\Sigma(\mb k)$,
where $\Sigma$ is the self energy and includes the Hartree and Fock
energies:

\begin{subequations}\label{eq:F_ALL}

\begin{eqnarray}
F & = & E_{0}+E_{B}-TS,\label{eq:F_FmNTS}\\
E_{0} & = & {\rm V}\int\frac{\mr{d^{3}}\mb k}{(2\pi)^{3}}\kappa(\mb k)G_{\mb k},\label{eq:F_E0}\\
E_{B} & = & \frac{{\rm V}}{2}\int\frac{\mr{d^{3}}\mb k}{(2\pi)^{3}}\int\frac{\mr{d^{3}}\mb{k'}}{(2\pi)^{3}}J_{\mb k}^{*}V(\mb k,\mb{k'})J_{\mb{k'}},\label{eq:F_Eb}\\
S & = & -{\rm V}\int\frac{\mr{d^{3}}\mb k}{(2\pi)^{3}}f_{\mb k}\log(f_{\mb k})+(1\!-\!f_{\mb k})\log(1\!-\!f_{\mb k}).\quad\label{eq:F_S}\\
J_{\mb k} & = & \frac{-\Delta(\mb k)}{2E(\mb k)}\tanh(\frac{E(\mb k)}{2T}),\label{eq:Jk_D}\\
G_{\mb k} & = & \frac{\kappa(\mb k)}{2E(\mb k)}\tanh(\frac{E(\mb k)}{2T})+\frac{1}{2}.\label{eq:Gk_D}\\
\Sigma(\mb k) & = & \int\frac{\mr d^{3}\mb q}{(2\pi)^{3}}\left\{ \bar{V}_{dd}(0)G(\mb q)-V(\mb k-\mb q)\mbox{G(\ensuremath{\mb q)}}\right\} ,
\end{eqnarray}

\end{subequations}

$E(\mb k)=\sqrt{\kappa(\mb k)^{2}+\Delta(\mb k)^{2}}$ is the full
quasi-particle energy, $f_{\mb k}=(1+e^{E(\mb k)/T})^{-1},$ V is
the volume, $V(\mb k)=\int e^{i\mb k\cdot\mb r}V_{\mr{dd}}^{\mr{(eff)}}(\mb r)\mr d^{3}\mb r$,
$V(\mb k,\mb{k'})=V(\mb k-\mb{k'})$, $\kappa(\mb k)=\xi(k)+\Sigma(\mb k)$,
$\xi(k)=\varepsilon_{0}(k)-\mu$, $\varepsilon_{0}(k)=\hbar^{2}k^{2}/2m$,
$G_{\mb k}=\left\langle \cop a_{\mb k}\aop a_{\mb k}\right\rangle $,
$J_{\mb k}=\left\langle \aop a_{\mb{-k}}\aop a_{\mb k}\right\rangle $,
and $f_{\mb k}=\left\langle \cop{\gamma}_{\mb k}\aop{\gamma}_{\mb k}\right\rangle $,
where $\cop a_{\mb k}$ and $\aop a_{\mb k}$ are the creation and
annihilation operators of particle in momentum eigenstate $\mb k$,
and $\aop{\gamma}$ is the quasiparticle operator satisfying $\left\langle \cop{\gamma}_{\mb k}\aop{\gamma}_{\mb k}\right\rangle =\delta_{\mb k,\mb k}$,
$\left\langle \aop{\gamma}_{\mb k}\aop{\gamma}_{\mb k}\right\rangle =0$,
and $\aop{\gamma}_{\mb k}=u_{\mb k}\cop a_{\mb k}+v_{\mb k}\aop a_{\mb k}$
for c-numbers $u_{\mb k}$ and $v_{\mb k}$.

Note that the BCS transition is a second order phase transition with
order parameter $\Delta$; therefore, it occurs at the point where
$\Delta=0$ goes from being a minimum of the free energy with respect
to $\Delta$ to a maximum or saddle point. Hence, only the Hessian
of $F$ with respect to $\Delta$ needs to be calculated. The transition
temperature is then simply the point where a negative eigenvalue occurs.

We will also need $V(\mb k)$ in spherical components:
\begin{align}
 & V(\mb k,\mb{k')}=\sum_{ll'm}Y_{l}^{m}(\hat{\mb k})V_{ll'}^{m}(k,k')Y_{l'}^{*m}(\hat{\mb{k'}}),\label{eq:Vkq_tospher-1}\\
 & V_{ll'}^{m}(k,k')=-\frac{(4\pi)^{2}}{\alpha}\sqrt{\frac{16\pi}{5}}i^{l'-l}(-1)^{m}I_{ll'}^{m}J_{ll'}(k,k'),\!\label{eq:Vllmqk-1}\\
 & J_{ll'}(k,k')=\int_{1}^{\infty}\!\frac{\mr dr}{r}j_{l}(kr)j_{l'}(k'r),\label{eq:Jkq-1}
\end{align}

\begin{align}
 & I_{ll'}^{m}=\sqrt{\frac{(2l+1)5(2l'+1)}{4\pi}}\begin{pmatrix}l & 2 & l'\\
0 & 0 & 0
\end{pmatrix}\begin{pmatrix}m & 0 & -m\\
l & 2 & l'
\end{pmatrix}.\label{eq:Illm-1}
\end{align}
Because we are dealing with identical p-wave fermions, we need only
consider odd angular momenta, $l$. Also, $V_{ll'}^{m}$ is non-zero
only when $l'=l$, or $l'=l\pm2$ \citep{2009-NJP_Bohn-Ticknor_DipScatUnivers}.
Notice that the dipolar interactions conserve the angular momentum
projection $m$. We will also need the self energy, which in a polarised
dipolar gas is anisotropic and, because the Hartree term is zero,
comes from the exchange interactions only; i.e., $-(2\pi)^{-3}\int\mr d^{3}\mb qV(\mb k-\mb q)\mbox{G(\ensuremath{\mb q)}}$.
We take $G\approx G_{0}=\theta(k_{F}-k)$, where $\theta$ is the
Heaviside theta function, which gives
\begin{align}
\Sigma(\mb k) & =\frac{-1}{2\pi}\left(1-3\cos^{2}(\theta_{k})\right)\sigma(k),\\
\sigma(k) & =\int_{1}^{\infty}\frac{1}{r^{4}}\left(\sin(k_{F}r)-k_{F}r\cos(k_{F}r)\right)j_{2}(kr)\mr dr.\label{eq:SelfEn_Fock_G0}
\end{align}
Notice that $\Sigma$ also conserves the angular momentum projection.
In order to solve the problem numerically, we must discretize the
radial $k$ direction and choose a maximum angular momentum ($l_{max}$).
We define a set of $n+1$ vertices, $x_{i}$, with $i\in[0,n]$, $x_{0}=0$,
and $x_{i}>x_{i-1}$, then we consider only the set of $\Delta(\mb k)$
functions that are constant when $k$ lies between vertices. We also
discretize $V(\mb k)$ on this grid. That is, we let\begin{subequations}\label{eq:vDelt_discrete_ALL}
\begin{align}
\Delta(\mb k) & \rightarrow\sum_{l=\mr{odd}}^{l_{max}}\sum_{m=0}^{l}\sum_{i=1}^{n}\tilde{\Delta}_{l}^{m}(i)\mathcal{U}_{x_{i-1}}^{x_{i}}(k)Y_{l}^{m}(\hat{\mb k}),\\
\bar{V}_{ll'}^{m}\left(i,j\right) & \equiv V_{ll'}^{m}\Bigl(\frac{x_{i-1}+x_{i}}{2},\frac{x_{j-1}+x_{j}}{2}\Bigr),\\
V_{ll'}^{m}(k,k') & \rightarrow\sum_{i,j=1}^{n}\mathcal{U}_{x_{i-1}}^{x_{i}}(k)\bar{V}_{ll'}^{m}\left(i,j\right)\mathcal{U}_{x_{j-1}}^{x_{j}}(k'),\\
\mathcal{U}_{x_{i-1}}^{x_{i}}(k) & \equiv\begin{cases}
1 & k\in[x_{i-1},x_{i}]\\
0 & \mr{otherwise}
\end{cases}.
\end{align}
\end{subequations}We now use these discrete forms and Eq.~(\ref{eq:SelfEn_Fock_G0})
to calculate the free energy, $F$. Then, we take the double derivative
of $F$ with respect to $\tilde{\Delta}$. Much cancellation occurs
giving the following surprisingly simple result for the discretized
Hessian matrix ($\mathcal{H}$) of $F$ with respect to the spherical
components of the gap.
\begin{multline}
\mathcal{\bar{H}}_{lmi}^{l'm'j}=\left.\frac{\partial^{2}F}{\partial\tilde{\Delta}_{l}^{m*}(i)\partial\tilde{\Delta}_{l'}^{m'}(j)}\right|_{\Delta=0}=\frac{\mr V\delta_{mm'}}{2(2\pi)^{3}}\mbox{\ensuremath{\times}}\\
\biggl\{\!K_{ll'}^{m}(i)\delta_{ij}+\!\frac{1}{(2\pi)^{3}}\sum_{\bar{l}\,\bar{\bar{l}}}K_{l\bar{l}}^{m}(i)V_{\bar{l}\,\bar{\bar{l}}}^{m}(i,j)K_{\bar{\bar{l}}l'}^{m}(j)\!\biggr\},\label{eq:Hessian}
\end{multline}
where the K matrix is given by
\begin{eqnarray}
K_{l,l'}^{m}(i) & \equiv & \int\mr d\hat{\mb k}K(i,\hat{\mb k})Y_{l}^{m*}(\hat{\mb k})Y_{l'}^{m}(\hat{\mb k}),\label{eq:Kmat_full}\\
K(i,\hat{\mb k}) & = & \int_{x_{i-1}}^{x_{i}}\mr dk\frac{k^{2}\tanh\left(\frac{\kappa(\mb k)}{2T}\right)}{2\kappa(\mb k)}.\label{eq:Kmat_radial}
\end{eqnarray}
Calculating whether the gas will be superfluid at a certain temperature
and density requires only the calculation of the $K$ and $V$ matrices
and then finding the lowest eigenvalue of the term inside the brackets
in Eq.~(\ref{eq:Hessian}). 

This method is applicable to any Hamiltonian for a homogeneous gas.
The only difficult numerics arise from calculating the the $K$ matrix
and a $\int_{1}^{\infty}\!\frac{\mr dr}{r}j_{l}(kr)j_{l'}(k'r)$ integral
that appears in $V_{l\,l'}^{m}$. The integral (with perhaps a different
power of $r$) is common to any potential, as is the whole $K$ matrix,
and it turns out that analytic solutions exist for both these functions
that are valid for most points on the grid. This means the Hessian
can be calculated easily and efficiently. Details are given in the
Appendix.

\section{Numerical Results\label{sec:Numerical-Results}}

\begin{figure}[H]
\noindent \centering{}\includegraphics[width=8.6cm]{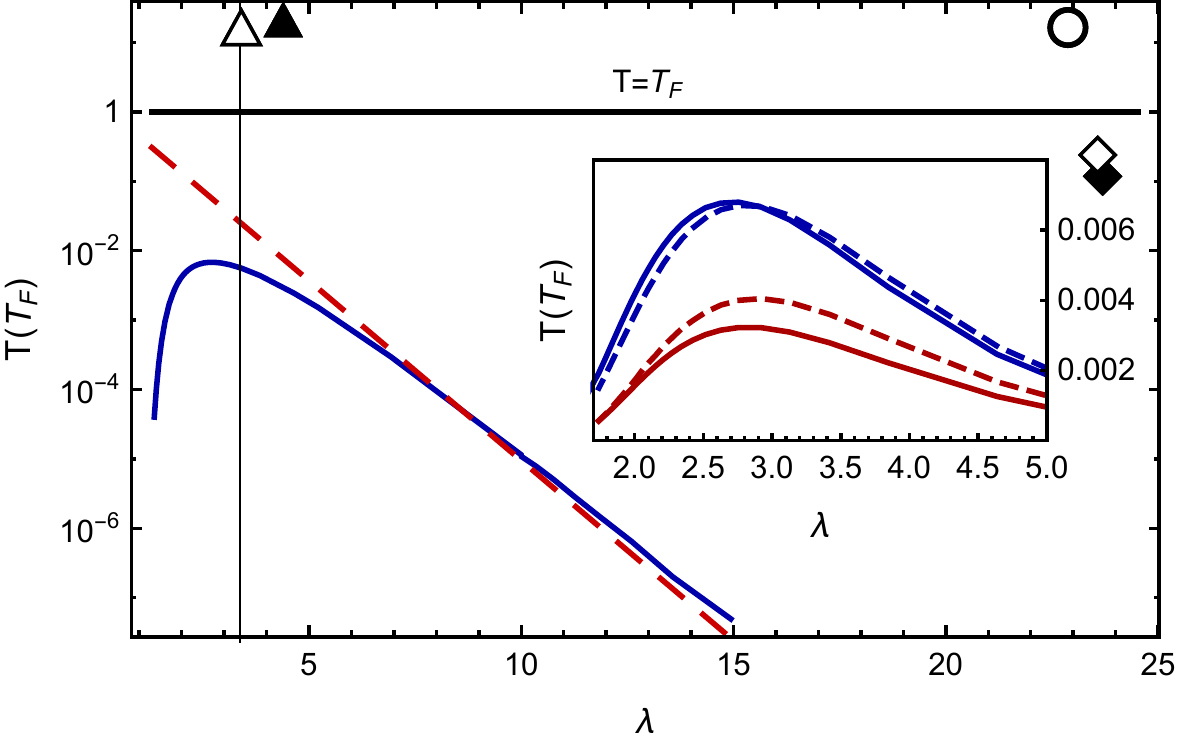}\protect\caption{Predictions for $T_{c}$ in a dipolar Fermi gas. $\lambda$ is the
dimensionless average distance between dipoles given by $(6\pi^{2})^{1/3}/(k_{F}\alpha C_{dd}m\hbar^{-2})$.
The straight line at the top is the position of the Fermi energy,
$T=T_{F}$. The solid blue line is our full numeric calculation. The
dashed red line is the theoretical prediction given in Ref.~\citep{2002-PRA_Baranov-Shly_BCSDipRenorm}.
The markers are the locations of current experiments: $\protect\mr{^{40}K^{87}Rb}$\citep{2010-Nature_JILA_UColdDipFermiGas}
at the theoretical maximum polarisation (\textifsymbol[ifgeo]{49}),
$\protect\mr{^{40}K^{87}Rb}$\citep{2010-Nature_JILA_UColdDipFermiGas}
at current experimental polarisations (\textbigcircle ) $\protect\mr{^{23}Na^{40}K}$\citep{2015-PRL_Harvard_NaKGstate}
at the theoretical maximum polarisation (\textifsymbol[ifgeo]{97}),
$\protect\mr{^{161}Dy}$\citep{2012-PRL_Lu-Burdick-Lev_DyFermiGas}
(\textifsymbol[ifgeo]{54}),$\protect\mr{^{167}Er}$ \citep{2014-PRL_Innsbruck_ErDeg}
(\textifsymbol[ifgeo]{102}), and $\protect\mr{^{53}Cr}$\citep{2015-PRA_CNRS_DegenCr}
is not plotted as it sits too far off to the right. Inset: $T_{c}$
vs $\lambda$ zoomed in at the peak in $T_{c}$. The bottom two lines
(Red) are with just $l=1$. The top two lines (blue) are with $l=1,3$.
The dashed lines are the results with exchange interactions switched
off. The upper solid (blue) line in the inset corresponds to the solid
(blue) line in the outer plot.\label{fig:Tc}}
\end{figure}

Results using this methodology for $T_{c}$, including the effect
of anisotropic exchange interactions, are shown in Fig.~\ref{fig:Tc}.
We use $l=1$ and $l=3$ contributions, $l=5$ makes almost no difference
to $T_{c}$. Also shown is the previous calculation from Ref. \citep{2002-PRA_Baranov-Shly_BCSDipRenorm}.
Our inclusion of $l=3$ states as well as exchange interactions gives
us a slightly higher $T_{c}$ at low densities. At higher densities,
the drop in $T_{c}$ reflects the fact that Ref. \citep{2002-PRA_Baranov-Shly_BCSDipRenorm}
is picking up contributions from p-wave dimers as discussed above.
The inset shows the effect of just considering $l=1$ and the effect
of turning off exchange interactions, which demonstrates that exchange
interactions can either increase or decrease $T_{c}$ depending on
the circumstances.

But what about the case of dipolar molecules that are chemically stable
against two-body collision\citep{2010-PRAR_Durham_Trimers}, such
as $\mr{^{23}Na^{40}K}$ \citep{2012-PRL_Harvard_NaK-AbsGS,2015-PRL_Harvard_NaKGstate},
or dipolar atoms? In a typical gas with short-range interactions,
three-body losses are proportional to the probability that two particles
will approach within a distance equal to the size of their bound state,
and that a third particle will venture within interactions range to
take away excess kinetic energy. For dipoles, we can see that for
inelastic scattering to occur, two dipoles must approach within each
others centrifugal barrier, but, due to long range interactions, the
third can absorb excess energy from a distance. Worse still, because
the gas interaction energy is not extensive, the whole gas can effectively
absorb excess energy from a collision pair simultaneously.

Referring back to Fig.~\ref{fig:Tc}, and comparing the filled and
dashed lines, we see that the cutoff only starts to affect $T_{c}$
around $\lambda\lesssim5$. At these densities, the average distance
between dipoles is only 2 to 30 times larger than the size of the
shallowest dimers (see the discussion in Fig.~\ref{fig:dip_boundstate}).
This is much smaller than is normal for typical dilute gasses with
short-range interactions. Given that once the molecules have tunneled
they can dimerise by expelling energy to multiple external dipoles
at once, and given their proximity to these other dipoles, it is not
unreasonable to expect dimerisation to occur at very high probability
inside the centrifugal barrier (This also agrees with exact numerical
calculations that give a three-body recombination rate of dipolar
fermions proportional to $C_{dd}^{8}$ \citep{2011-PRL_WangIncGreene_3BodyDipFermion}).
If we combine this with the possibility of long lived scattering chain
complexes \citep{2013-PRA_JILA-Que-Bohn_ScatChainComplex}, and of
inelastic spin-flip interactions in atoms\citep{2012-PRL_Lu-Burdick-Lev_DyFermiGas},
then to us it seems reasonable that our effective potential should
be valid also for chemically stable molecules and dipolar atoms as
well (for dipolar atoms, the bound states in Fig.~\ref{fig:dip_boundstate}
just represent the shallowest rovibrational states of a molecule).

\section{Discussion\label{sec:Conclusion}}

\subsection*{Comparison with other work}

It is worthwhile to now compare our results to those of Ref.~\citep{2002-PRA_Baranov-Shly_BCSDipRenorm}
to understand where and how they differ. In Ref.~\citep{2002-PRA_Baranov-Shly_BCSDipRenorm},
the same calculation was performed, except using the renormalisation
method from Refs. \citep{1989-PRL_Randiera,1990-PRB_Randiera}. That
is, instead of finding non-zero solutions to the gap equation,
\begin{equation}
\Delta(\mb k)=-\int\frac{\mr{d^{3}}\mb q}{(2\pi)^{3}}V(\mb k,\mb{q)}\frac{\Delta(\mb q)}{2E(\mb q)}\mr{tanh}\Bigl(\frac{E(\mb q)}{2T}\Bigr),\label{eq:gapeq}
\end{equation}

one solves the renormalised gap equation,

\begin{equation}
\Delta(\mb k)=\int\frac{\mr{d^{3}}\mb q}{(2\pi)^{3}}T(\mb k,\mb q)\left\{ \frac{1}{2\varepsilon_{0}(k)}-\frac{\mr{tanh}(\frac{E(\mb q)}{2T})}{2E(\mb q)}\right\} \Delta(\mb q),\label{eq:gapeq_renorm}
\end{equation}

which is the same equation, but in terms of the scattering T-matrix,
$T$, rather than the potential, $V$. In order to solve this equation
the authors first make the approximation $T=V$ (the born approximation),
and then to first order (small $k$) remove the $\frac{1}{2\varepsilon_{0}(k)}$
term. They are therefore solving almost the same equations as we do,
but with a $V$ that is not cut off at $r_{b}$. However, they use
the approximation that $\Delta$ is concentrated around the Fermi
surface, therefore making the region around $k_{F}\sim r_{b}^{-1}$
irrelevant for small densities or small interaction strengths (i.e.,
large $\lambda$). It is therefore not surprising that in Fig.~\ref{fig:Tc}
both solutions agree closely at large values of $\lambda$.

As $\lambda\rightarrow1$, however, the cutoff becomes very relevant,
and the two solutions diverge. Although we have justified our methodology
in detail here, one might be tempted to argue that the method presented
in Ref. \citep{2002-PRA_Baranov-Shly_BCSDipRenorm} has a physical
basis in the fact that it does not depend on any cutoff, whereas our
solutions are highly dependent on the choice of $r_{b}$ for high
densities. Below we elucidate how the renomalisation method presented
in Ref. {[}1{]} does implicitly introduce a short range cut-off.

Such an assertion is based on the idea that the term in brackets in
Eq.~(\ref{eq:gapeq_renorm}) approaches zero for large $k$, which
makes the details of $T$ irrelevant at large $k$, and therefore
``renormalizes'' out the short range behavior the gas. However,
for the case of a dipolar potential considered here, things are more
complicated. First notice that due to its $r^{-3}$ behavior, a bare
dipolar potential does not form a well-defined Hamiltonian. That is,
if we use a cutoff, as that cutoff approaches zero, the number of
negative energy eigenstates approaches infinity and the energy of
the deepest eigenstates approaches negative infinity. The T-matrix
for a bare dipolar interaction is therefore also not defined without
also choosing a cutoff.

For identical fermions, however, this difficulty is somewhat allayed
by the presence of the centrifugal barrier. Because identical fermions
with low scattering energy can not approach closely, $T$, for long
wavelengths, is mostly independent of the cutoff. This is referred
to as quasi-universal dipolar scattering \citep{2009-NJP_Bohn-Ticknor_DipScatUnivers}.
Also, we can easily calculate this universal small $k$ behavior of
$T$ by noting that, because of this insensitivity, we can put the
cutoff anywhere within the centrifugal barrier and still get the same
answer (provided the cutoff doesn't put the system on resonance).
If we choose the cutoff to be at $r_{b}$ and choose $V$ to be zero
everywhere within that cutoff, it is easy to see that $V$ is small
everywhere, and we can therefore use the born approximation $T\approx V$.
This is exactly what is done in Ref.~\citep{2002-PRA_Baranov-Shly_BCSDipRenorm}.
It is important to realise that in choosing the born approximation
in this way, a cutoff is implicitly chosen. For example, as we move
further inside the centrifugal barrier of a dipolar potential, the
strength of the potential is much stronger, and because of this, to
correctly describe $T$ in this case, one would have to use the second
order born approximation. Such a choice for $T$ would be equivalent
to using a different implicit value for the cutoff. 

For large values of $\lambda$ our solutions are insensitive to the
choice of $r_{b}$ just as the solutions of reference Ref.~\citep{2002-PRA_Baranov-Shly_BCSDipRenorm}
are insensitive to the approximation used to calculate $T$. As $\lambda$
gets smaller, both our results, and those of Ref.~\citep{2002-PRA_Baranov-Shly_BCSDipRenorm}
become highly dependent on this choice. 

At this point one should note that, although the method used in Ref.~\citep{2002-PRA_Baranov-Shly_BCSDipRenorm}
is equivalent to choosing a cutoff, it is not equivalent to making
the potential zero inside this cutoff. More specifically, they are
effectively choosing an effective potential, $V'$, such that the
T-matrix generated by $V'$ is equal to the bare dipolar potential.
This $V'$ agrees with our $V_{\mr{dd}}^{\mr{(eff)}}$, Eq.~(\ref{eq:Veff_r}),
outside $r_{b}$ but not inside. There is no reason a priori why either
choice for the effective potential would be more correct. In this
work we argued that it is unreasonable to expect the region inside
the centrifugal barrier to contribute to the superfluid, and the optimal
choice is zero.

\subsection*{The $k_{F}\sim1/r_{b}$ regime}

In order for a superfluid to appear, the thermalisation rate into
bound Cooper pairs must be faster than the rate of quenching into
tightly bound pairs and trap loss. Although most of this work is dedicated
to dealing with the effect of this quenching, we have not explicitly
calculated any transition rates. Rather, we have noted that, within
certain temperature-density regimes, a transition to a BCS state cannot
occur faster than the quenching rate because \emph{the BCS state itself
is the unstable tightly bound pairs}. We have therefore calculated
a temperature upper bound for which a BCS is possible \emph{provided}
thermalisation could occur fast enough.

The $k_{F}\sim1/r_{b}$ region is of interest because it is the quantum
degenerate regime where the kinetic energy and dipolar energy are
of comparable order. The results here show that a homogeneous dipolar
gas should be unstable in this regime and hence $T_{c}$ goes to zero
in  Fig.~\ref{fig:dip_boundstate}. However, a number of experimental
methods exist which could be used to artificially stabilize the gas.
In particular, the method presented in Refs.~ \citep{2007-PRA_Micheli-Zoller_InterTaylor,2007-PRL_Buchler-Zoller_InterTaylor},
or a bilayer trap discussed in Ref.~\citep{2011-PRA_BaranovZoller_BilayerDipFerm},
could both be used to overcome the problems discussed here.

\subsection*{Resonant scattering}

All of the results discussed here assume that the dipolar potential
is off resonance. In the case of resonant scattering, experiments
have already shown that the transition temperature can be made as
high as $0.2T_{c}$. This is much higher than the results given here
for BCS superfluidity due to the long-range part of the potential.
It is clear then that for the case of resonant scattering, any contribution
from the long-range part will be overwhelmed by the resonant scattering,
and the system should behave like a system of short-range, p-wave-interacting
atoms. As discussed in Section \ref{sec:dip-bond}, these systems
have already been shown to be unstable, and there is no reason to
believe a dipolar gas would be any different given the insignificance
of the long range contribution in the resonant regime.

\subsection*{Conclusion}

The first purpose of this work is to show that previous work on dipolar
Fermi gasses have calculated a transition to a tightly bound BEC pair
and not a BCS superfluid. We pointed out the inherit instability of
these pairs and investigated the effect this phenomena has on the
transition temperature. The second purpose of this work is to present
a general numerical method for calculating $T_{c}$ for systems where
the particles have complicated self energy configurations. It is particularly
well suited to the dipolar gas problem because of the anisotropy of
the Fermi surface, and it is also applicable more generally to systems
with complicated self energies.

\appendix

\section*{Appendix: Analytic and numerical techniques}

\subsection*{A1) The $K$ matrix integral}

The $K$ matrix (Eqs.~(\ref{eq:Kmat_full}) and (\ref{eq:Kmat_radial}))
typically has a few hundred elements corresponding to each different
basis point. For each basis point, the integral in Eqs.~(\ref{eq:Kmat_full})
and (\ref{eq:Kmat_radial}) must be performed. Furthermore, every
time $\varrho$ or $\tau$ is changed, every basis point must be recalculated.
For the case of an isotropic $V(\mb k)$, the integral is not overly
difficult because the angular part disappears. For the anisotropic
case however, each grid point requires that a three-dimensional integral
be performed (although in the case of dipoles, cylindrical symmetry
removes one of the angular dimensions, leaving a two-dimensional integral).
Also, the fact the potential is anisotropic means that the cross terms
of different $l$ and $l'$ become non zero, leading to still more
bases to calculate. All this is compounded by the fact that the integrand
is very tightly peaked near the Fermi surface, especially at low temperatures.
For these reasons, the integral is too challenging to be done by brute
force numerical methods. Fortunately, analytic solutions exist for
the radial part of the integral, which is the most time consuming
because it contains the peak at the Fermi surface.

To calculate $K(i,\hat{\mb k})\equiv\int_{x_{i-1}}^{x_{i}}\mr dk\frac{k^{2}\tanh\left(\frac{\kappa(\mb k)}{2T}\right)}{2\kappa(\mb k)}$,
we can expand $\Sigma(\hat{\mb k},k)$ in a power series to order
$k^{2}$. This gives $\kappa=\alpha(\hat{\mb k})k^{2}+\beta(\hat{\mb k})k+\omega(\hat{\mb k})$.
We then get the following asymptotic formula for the indefinite integral:

\begin{multline}
\int\mr dk\frac{k^{2}}{2(\alpha k^{2}+\beta k+\omega)}\tanh\left(\frac{\alpha k^{2}+\beta k+\omega}{2T}\right)\\
\sim\frac{\sqrt{2T}}{4\alpha^{3/2}}A-\frac{\beta}{4\alpha^{2}}B+\frac{\beta^{2}}{16\alpha^{5/2}\sqrt{2T}}C,\label{eq:asym-indef-K-int}
\end{multline}
where,
\begin{align}
A & =\tanh\left(\frac{\nu^{2}-\nu_{F}^{2}}{4T}\right)\frac{1}{\sqrt{T}}\biggl\{\nu+\frac{\nu_{F}}{2}\log\left(\frac{\nu_{F}-\nu}{\nu_{F}+\nu}\right)\nonumber \\
 & -i\pi\frac{\nu_{F}}{4}\biggr\}+\theta(\nu^{2}-\nu_{F}^{2})\frac{\nu_{F}}{\sqrt{T}}\biggl\{-2+\gamma+\log\left|\frac{4\nu_{F}^{2}}{\pi T}\right|\nonumber \\
 & +\frac{(\pi T)^{2}}{12\nu_{F}^{4}}+\frac{7(\pi T)^{4}}{96\nu_{F}^{8}}+O\biggl(\frac{T{}^{6}}{\nu_{F}^{12}}\biggr)\biggr\},
\end{align}
\begin{align}
B & =\tanh\left(\frac{\nu^{2}-\nu_{F}^{2}}{4T}\right)\left\{ \log\left(\frac{\nu^{2}-\nu_{F}^{2}}{4T}\right)-i\frac{\pi}{2}\right\} \nonumber \\
 & +\theta(\nu^{2}-\nu_{F}^{2})2\left\{ \gamma-\log\left|\frac{\pi}{4}\right|\right\} ,
\end{align}
\begin{align}
C & =\tanh\left(\frac{\nu^{2}-\nu_{F}^{2}}{4T}\right)\sqrt{T}\biggl\{\frac{2}{\nu_{F}}\log\left(\frac{\nu_{F}-\nu}{\nu_{F}+\nu}\right)\nonumber \\
 & -i\frac{\pi}{\nu_{F}}\biggr\}+\theta(\nu^{2}-\nu_{F}^{2})\frac{4\sqrt{T}}{\nu_{F}}\biggl\{\gamma-\log\left|\frac{\pi T}{4\nu_{F}^{2}}\right|\nonumber \\
 & -\frac{(\pi T)^{2}}{4\nu_{F}^{4}}-\frac{49(\pi T)^{4}}{96\nu_{F}^{8}}-O\biggl(\frac{T{}^{6}}{\nu_{F}^{12}}\biggr)\biggr\},
\end{align}

and we use the definitions $\nu=\sqrt{2\alpha}k+\frac{\beta}{\sqrt{2\alpha}}$,
$\eta=\frac{\beta^{2}}{4\alpha}-\omega$, and $\nu_{F}=\pm\sqrt{2\eta}$.
This formula is an asymptotic solution for the indefinite integral
in the limit $\left|\kappa(k)\right|\rightarrow\infty$. It is valid
so long as $\kappa$ is either strictly increasing or strictly decreasing
on the integration region. It converges very quickly, and for $\left|\kappa(k)/2T\right|\geq12$
it is exact to 10 decimal places. The correct solution for $\nu_{F}$
depends on $\kappa$ and $\alpha$. If $\kappa$ is strictly increasing
and $\alpha>0$, or if $\kappa$ is strictly decreasing and $\alpha<0$
then one must take the positive solution, otherwise one needs the
negative solution. If one requires more accuracy, the expansion must
be carried out beyond $O(T{}^{6}/\nu_{F}^{12})$. If an end point
of one of the grid points lies too close to the Fermi surface, then
the grid point can either be moved, or the radial integral can be
done numerically for that particular point only.

This formula can be derived by changing the integration variable to
$\kappa$ and using integration by parts to get a $\mr{sech}^{2}$
in the integral instead of tanh. Because $\mr{sech}^{2}$ decays exponentially
quickly, we can replace the integration limits from $[\kappa(x_{i}),\kappa(x_{i+1})]$
to $[-\infty,\infty]$ and perform the remaining integral using standard
integrals.

With this analytic solution for the radial part of the integral, the
angular part can be done numerically in a short amount of time.

\subsection*{A2) The $J=\int_{1}^{\infty}\!\frac{\protect\mr dr}{r}j_{l}(kr)j_{l'}(k'r)$
integral}

Whenever one expands $V(\mb k)$ into its radial components, they
will end up with an integral of the form of $J$ in Eq.~(\ref{eq:Jkq-1}).
Depending on the form of the potential, there may be a different power
of $r$ in the integral. This integral needs to be calculated once
for each element in the $V$ matrix. $V$ contains a number of elements
on the order of the number of basis squared. In our case, to calculate
the $V$ matrix, the $J$ integral had to be calculated over 40,000
times. For this reason, the integral must be calculated very quickly
as well as accurately. The following method is sufficient for this
task.

For $k\approx k'$, the integral must be performed as follows. First
note that the Bessel functions converge slowly as $r\rightarrow\infty$
and are oscillatory. This makes straight numerical integration very
difficult. However, we should recall the asymptotic properties of
the Bessel functions
\begin{align}
j_{l}(kr) & \rightarrow j_{l}^{\infty}(kr)\quad\mr{as}\quad r\rightarrow\infty,
\end{align}
where
\begin{equation}
j_{l}^{\infty}(kr)=\frac{1}{kr}\cos\left(kr-(l+1)\frac{\pi}{2}\right).
\end{equation}
If we define
\begin{align}
J_{l,l'}^{\infty}(k,q) & \equiv\int_{1}^{\infty}\frac{1}{r}j_{l}^{\infty}(kr)j_{l'}^{\infty}(qr)\mr dr\\
\widetilde{J}_{l,l'}(k,q) & \equiv\int_{1}^{\infty}\frac{1}{r}\left(j_{l}(kr)j_{l'}(qr)-j_{l}^{\infty}(kr)j_{l'}^{\infty}(qr)\right)\mr dr
\end{align}
then the integrand in $\widetilde{J}$ has the oscillatory asymptotic
part removed, making numerical integration much easier, and $J_{l,l'}^{\infty}(k,q)$
can be calculated analytically using standard techniques. The full
integral is then
\begin{equation}
J_{l,l'}(k,q)=\widetilde{J}_{l,l'}(k,q)+J_{l,l'}^{\infty}(k,q).
\end{equation}

The above technique only removes the oscillatory part of $j$ to first
order. When we have $k\ll q$ or $q\ll k$, the integrand becomes
extremely oscillatory even with the $j^{\infty}$ terms removed. The
numerical integration becomes so time consuming that calculating the
integral even once is difficult, let alone thousands of times. Fortunately,
we can expand $J_{l,l'}(k,q)$ in a power series expansion that is
valid for $k\ll q$ or $q\ll k$. Here we will assume $k\ll q$.

Let $\rho_{l}^{(n)}(kr)$ be the n'th order series expansion of $j_{l}(kr)$
about $k=0$. Then define the indefinite integral

\begin{equation}
\mathscr{J}_{l,l'}^{(n)}(k,q;r)\equiv\int\frac{1}{r}\rho_{l}^{(n)}(kr)j_{l'}(qr)\mr dr.
\end{equation}

$\mathscr{J}_{l,l'}^{(n)}(k,q;r)$ is a divergent expansion for the
function $\int\frac{1}{r}j_{l}(kr)j_{l'}(qr)\mr dr$. For small $r$
the two functions agree, but no matter how many terms in the expansion
of $\rho_{l}^{(n)}(kr)$ one uses, the function $\mathscr{J}_{l,l'}^{(n)}(k,q;r)$
will always diverge for large $r$, and therefore one has to be careful
about attempting erroneous actions such as $J_{l,l'}(k,q)=\mathscr{J}_{l,l'}^{(n)}(k,q;\infty)-\mathscr{J}_{l,l'}^{(n)}(k,q;0).$

However, note the following. The integrand $\frac{1}{r}j_{l}(kr)j_{l'}(qr)$
converges to zero as $r$ increases. This means that the approximate
integrand $\frac{1}{r}\rho_{l}^{(n)}(kr)j_{l'}(qr)$ also converges
to zero before it blows up at larger values of $r$. Let $r_{max}^{(n)}$
be the value of $r$ at which $\frac{1}{r}\rho_{l}^{(n)}(kr)j_{l'}(qr)$
is closest to zero and still a good approximation for $\frac{1}{r}j_{l}(kr)j_{l'}(qr)$.
If $\frac{1}{r}j_{l}(kr_{max})j_{l'}(qr_{max})$ is sufficiently small
then

\begin{align}
J_{l,l'}(k,q) & \approx\int_{1}^{r_{max}}\frac{1}{r}j_{l}(kr)j_{l'}(qr)\mr dr\nonumber \\
 & \approx\mathscr{J}_{l,l'}^{(n)}(k,q;r_{max})-\mathscr{J}_{l,l'}^{(n)}(k,q;0)\nonumber \\
 & \approx-\mathscr{J}_{l,l'}^{(n)}(k,q;0).
\end{align}

Now it turns out that the smaller $k$ is, the faster the integrand
converges, and the less terms (smaller $n$) one needs to consider
in $\rho_{l}^{(n)}(kr)$. So, it in fact turns out that $-\mathscr{J}_{l,l'}^{(n)}(k,q;0)$
is \emph{the} series expansion of $J_{l,l'}(k,q)$ around small $k$.
We can likewise do the same for small $q$.

Finally, it is important to note that $J_{l,l'}(k,q)$ is not analytic
at $k=q$. This means that the small $k$ and small $q$ expansions
can only work up to a point. Once we are in the $k\approx q$ regime
we must resort to the first method described above. 

\bibliographystyle{apsrev4-1}
\bibliography{DipFermGas,FermGas,ThesisOther}

\end{document}